\documentclass[aps,prl,twocolumn,superscriptaddress,floatfix,preprintnumbers,nofootinbib]{revtex4-2}
\usepackage{amscd,
amsfonts,
amsmath,
amsthm, 
amssymb, 
bm, 
braket, 
comment, 
eso-pic, 
esvect, 
eucal,
graphicx, 
here,
hyperref,
latexsym, 
multirow, 
ragged2e, 
siunitx, 
slashed, 
xcolor, 
}
\usepackage{afterpage}
\usepackage{changepage}
\usepackage{soul}
\usepackage{float}
\usepackage{mathrsfs}
\usepackage{wrapfig,caption,subcaption}
\usepackage{colortbl}
\usepackage{ascmac}
\usepackage{subfiles}
\usepackage{cleveref}
\usepackage{mathtools}
\usepackage[normalem]{ulem}
\hypersetup{
    colorlinks=true,
    linkcolor=blue,
    citecolor=magenta,      
    urlcolor=cyan}
\usepackage[english]{babel}
\usepackage{lipsum}

\newcounter{num}

\begin{document}
\preprint{MI-HET-858, MITP-25-042}
\title{Opening up New Parameter Space for Sterile Neutrino Dark Matter}
\author{P.~S.~Bhupal Dev}
\email{bdev@wustl.edu}
\affiliation{Department of Physics and McDonnell Center for the Space Sciences,  
Washington University, Saint Louis, MO 63130, USA}
\affiliation{PRISMA$^+$ Cluster of Excellence \& Mainz Institute for Theoretical Physics, 
Johannes Gutenberg-Universit\"{a}t Mainz, 55099 Mainz, Germany}
\author{Bhaskar Dutta}
\email{dutta@tamu.edu}
\affiliation{Mitchell Institute for Fundamental Physics and Astronomy, Department of Physics and Astronomy, Texas A\&M University, College Station, TX 77843, USA}

\author{Srubabati Goswami}
\email{sruba@prl.res.in}
\affiliation{Theoretical Physics Division, Physical Research Laboratory,  
Navrangpura, Ahmedabad 380009, India}

\author{Jianrong Paul Tang}
\email{jianrong.t@wustl.edu}
\affiliation{Department of Physics, 
Washington University, Saint Louis, MO 63130, USA}

\author{Aaroodd Ujjayini Ramachandran}
\email{aaroodd@wustl.edu}
\affiliation{Department of Physics, 
Washington University, Saint Louis, MO 63130, USA}

\begin{abstract}
Sterile neutrinos are compelling dark matter (DM) candidates, yet the minimal production mechanism solely based on active ($\nu_a$)-sterile ($\nu_s$) oscillations is excluded by astrophysical observations. Non-standard self-interactions in either active ($\nu_a-\nu_a$) or sterile ($\nu_s-\nu_s$) sector are known to alter the sterile neutrino DM  production in the early Universe, which could alleviate the tension with astrophysical constraints to some extent. Here we propose a novel solution where scalar-mediated non-standard interactions between active and sterile neutrinos ($\nu_a-\nu_s$) generate new production channels for $\nu_s$,  independent of the active-sterile mixing and without the need for any fine-tuned resonance or primordial lepton asymmetry. This framework enables efficient sterile neutrino DM production even at vanishingly small mixing angles and opens up new viable regions of parameter space that can be tested with future $X$-ray and gamma-ray observations. It also provides a new mechanism for observable neutrino-DM interactions while being consistent with DM relic density and has broad phenomenological implications for astrophysical neutrinos and beyond.    
\end{abstract}
\maketitle


\textbf{\textit{Introduction.}}---
\label{sec:introduction}
The nature of dark matter (DM) remains an open question of fundamental importance in physics. As the so-called `WIMP miracle' with GeV--TeV scale thermal freeze-out DM continues to lose its appeal due to null experimental searches over the past several decades~\cite{Arcadi:2024ukq}, there is a growing interest in alternatives to the WIMP paradigm~\cite{Baer:2014eja}, such as WIMPless DM~\cite{Feng:2008ya}, freeze-in DM~\cite{Hall:2009bx}, axions~\cite{Adams:2022pbo}, sterile neutrinos~\cite{Abazajian:2017tcc}, wave-like DM~\cite{Hui:2021tkt} and primordial black holes~\cite{Bird:2022wvk}. Here we focus on keV-scale sterile neutrinos, which are particularly well-motivated candidates for (warm) DM due to their potential connection to other outstanding puzzles like neutrino mass and matter-antimatter asymmetry~\cite{Asaka:2005an, Drewes:2016upu,Boyarsky:2018tvu, Dasgupta:2021ies}.

\begin{figure}[t!]
    \includegraphics[width=0.48\textwidth]{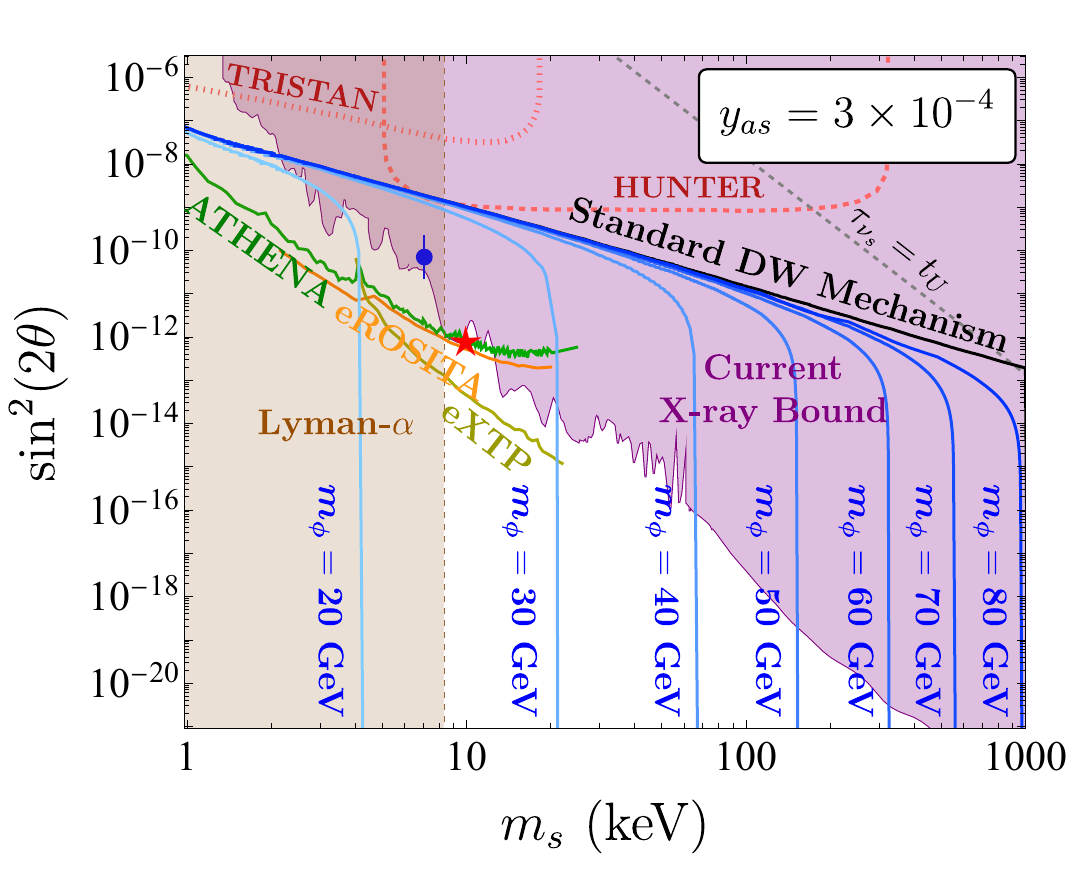}
    \caption{\justifying Sterile neutrino mass versus mixing angle for a fixed active-sterile neutrino coupling $y_{as}=3 \times 10^{-4}$. Shaded regions indicate astrophysical constraints, while the open curves on the left represent future $X$-ray sensitivities. Red dotted and dashed curves show projected $\beta$-decay experiment sensitivities. The gray dashed line marks the sterile neutrino lifetime equal to the age of the Universe. The blue contours satisfy the correct DM relic abundance $\Omega_s h^2 = 0.12$ for different values of the mediator mass $m_\phi$. The black line corresponds to the correct relic density using the DW mechanism. The red star marks a benchmark point (BP) used later to illustrate the yield behavior. The blue point corresponds to the purported 3.5 keV $X$-ray line signal.}
    \label{fig:param_space}
\end{figure}

In a minimal extension of the Standard Model (SM), sterile neutrinos are produced non-thermally in the early universe through active ($\nu_a$)-sterile ($\nu_s$) neutrino oscillations. This non-resonant production, known as the \textit{Dodelson-Widrow (DW) mechanism}~\cite{Dodelson:1993je}, is theoretically attractive since it relies solely on neutrino mixing and requires no new particles beyond sterile neutrinos. However, achieving the correct DM relic abundance in this scenario requires relatively `large' mixing angles that are strongly constrained by observational data, mostly from $X$-ray and gamma-ray line searches for the radiative decay $\nu_s\to \nu_a+\gamma$~\cite{Borriello:2011un,Riemer-Sorensen:2014yda, Anderson:2014tza,Hofmann:2016urz, Dessert:2018qih, Ng:2019gch, Calore:2022pks, Krivonos:2024yvm, Yin:2025xad}, effectively excluding the entire DW parameter space; see Fig.~\ref{fig:param_space}. 

A possible alternative to the DW mechanism is the \textit{Shi-Fuller mechanism}~\cite{Shi:1998km}, in which a primordial lepton asymmetry ($\eta_L \sim 10^{-3} \text{ to } 10^{-1}$) resonantly enhances the sterile neutrino production. However, the required lepton asymmetry to yield efficient DM production is several orders of magnitude larger than the observed baryon asymmetry (\( \eta_B \sim 10^{-10} \)), challenging its consistency with standard cosmology and Big Bang Nucleosynthesis (BBN)~\cite{Shi:1993hm, Boyarsky:2009ix, Bodeker:2020hbo}, although flavor-specific lepton asymmetries can mitigate the BBN bounds to some extent~\cite{Gorbunov:2025nqs}.

These challenges motivate the need for alternative sterile neutrino DM  production mechanisms,  involving non-standard interactions (NSI) in the neutrino sector~\cite{Dev:2019qno} or using heavy particle decays~\cite{Datta:2021elq,Abada:2023mib, Abada:2025gvc}. Prior studies using NSI have explored the effect of self-interactions between the active neutrinos ($\nu_a-\nu_a$)~\cite{DeGouvea:2019wpf,Kelly:2020pcy, Alonso-Alvarez:2021pgy, Benso:2021hhh} or between the sterile neutrinos ($\nu_s-\nu_s$)~\cite{Johns:2019cwc, Bringmann:2022aim, Astros:2023xhe} on the DM production. In this Letter, we explore an intriguing new possibility, i.e. NSI between active and sterile neutrinos ($\nu_a-\nu_s$). While active-sterile NSI has been previously explored in the contexts of BBN~\cite{Babu:1991at} and high-energy neutrinos~\cite{Fiorillo:2020jvy, Fiorillo:2020zzj}, its role in sterile neutrino DM production has not been studied before.

As summarized in Fig.~\ref{fig:param_space}, active-sterile neutrino interactions lead to unique features very different from the active-active and sterile-sterile self-interaction scenarios. In particular, the active-sterile NSI introduce new number-changing processes, such as $\nu_a \nu_a \leftrightarrow \nu_s \nu_s $, providing an efficient sterile neutrino production channel that differs fundamentally from the DW mechanism, as it does not rely on the active-sterile mixing. As a result, it  extends the accessible parameter space to arbitrarily small mixing angles. It significantly improves upon earlier self-interaction-based production scenarios, which necessarily depend on the mixing and underproduce sterile neutrinos below a threshold mixing angle. Additionally, unlike scenarios involving sterile self-interactions, this mechanism does not rely on any fine-tuned resonance to achieve efficient DM production. 

\medskip
\textbf{\textit{Model}}---
\label{sec:model}
We consider a keV-scale Majorana sterile neutrino, which  mixes  with active neutrinos with strength  $\theta$ and also participates in NSI via a complex scalar mediator, assumed to have mass \( m_\phi \gtrsim \mathcal{O}(\mathrm{GeV}) \). 
The relevant effective interaction Lagrangian is given by  
\begin{align}  
    -\mathcal{L} \supset y_{as} \bar{\nu}_a \nu_s \phi +{\rm h.c.} \, ,
    \label{eq:Leff}
\end{align} 
where we have taken the coupling constant $y_{as}$ to be real.\footnote{A pseudoscalar mediator with an imaginary coupling will lead to the same physical effects in the relativistic limit, as shown in Ref.~\cite{Fiorillo:2020zzj} for the high-energy neutrino case.} We have also suppressed the active neutrino flavor index, as the DM production happens above the neutrino decoupling and all flavors lead to qualitatively similar results. 
A concrete ultraviolet-completion of the effective Lagrangian~\eqref{eq:Leff} in terms of a CP-conserving two-Higgs-doublet model plus a complex
scalar singlet can be found in Ref.~\cite{Dutta:2020scq}. Note that the $\phi$ field does not acquire a vacuum expectation value, and therefore, there is no induced active-sterile neutrino mixing due to the interaction~\eqref{eq:Leff}. 

We assume no primordial population of sterile neutrinos, which are instead produced entirely from the active sector. For a GeV-scale scalar mediator, the decay rate of $\phi\to \nu_a\nu_s$ is much faster than the relevant production timescales at $\mathcal{O}({\rm MeV})$, provided the coupling satisfies $y_{as} \gtrsim 10^{-8}$. Under this condition, the evolution of the mediator $\phi$ can be safely neglected.

Since sterile neutrinos are not in thermal equilibrium with the active sector, we model their distribution using a modified Fermi--Dirac (FD) form:
\begin{align}
    f_s(p) = \frac{\alpha }{e^{p/T_s} + 1} \, ,
    \label{eq:dist}
\end{align}
where $p$ is the magnitude of the 3-momentum, \( T_s \) is an effective temperature and \( 0 < \alpha \lesssim 1 \) is a normalization factor encoding the departure from thermal equilibrium. 
Both \( T_s \) and \( \alpha \) are treated as functions of the SM temperature \( T \). In the dilute regime \( f_s(p) \ll 1 \), where quantum statistical effects are negligible, Eq.~\eqref{eq:dist} approaches the Maxwell--Boltzmann limit. It also recovers the fully thermalized case in the `large' coupling limit (\( \alpha = 1, T_s = T \)) and coincides with the active neutrino distribution $f_a(p)=1/(e^{p/T}+1)$.

\begin{figure}[htbp!]
    \centering
    \begin{minipage}[c]{0.19\textwidth}
        \centering
        \includegraphics[width=\textwidth]{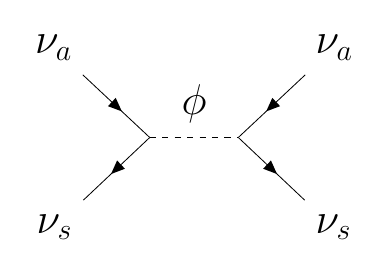}\subcaption{}
         \label{fig:scattering1}
        \includegraphics[width=\textwidth]{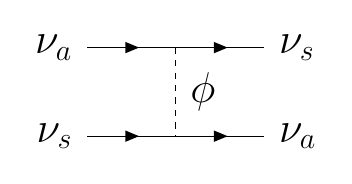}\subcaption{}
          \label{fig:scattering2}
    \end{minipage} \hspace{0.015\textwidth}
     \begin{minipage}[c]{0.19\textwidth}
        \centering
        \includegraphics[width=\textwidth]{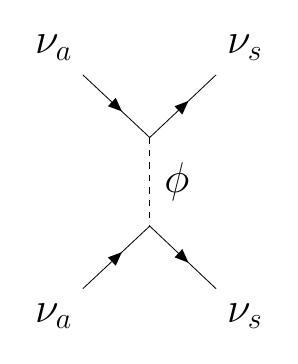}
        \subcaption{}
          \label{fig:numberchanging}
    \end{minipage} \hspace{0.015\textwidth}
\begin{minipage}[c]{0.19\textwidth}
    \centering
    \includegraphics[width=\textwidth]{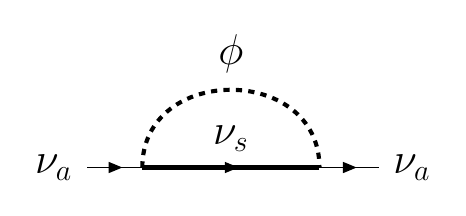}\subcaption{}
      \label{fig:selfenergy1}
\end{minipage} \hspace{0.015\textwidth}
\begin{minipage}[c]{0.19\textwidth}
    \centering
    \includegraphics[width=\textwidth]{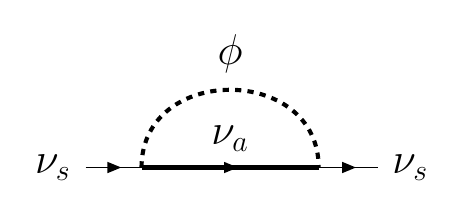}\subcaption{}
    \label{fig:selfenergy2}
\end{minipage}
   \caption{\justifying
Feynman diagrams illustrating active–sterile NSI contributions to scattering, number-changing interactions, and thermal potentials.}
    \label{fig:feynman_diagrams}
\end{figure}

The new NSI~\eqref{eq:Leff} opens up additional channels for active-sterile neutrino scattering (see Figs.~\ref{fig:scattering1},~\ref{fig:scattering2})  and introduces new contributions to the neutrino self-energy (see Figs.~\ref{fig:selfenergy1},~\ref{fig:selfenergy2}). It also enables a number-changing process, $\nu_a \nu_a \to \nu_s \nu_s$ (see Fig.~\ref{fig:numberchanging}), which plays a central role in the production of sterile neutrinos.  Unlike the DW mechanism, this does not rely on active–sterile mixing and can remain efficient even when mixing angles vanish. These reactions are especially important at high temperatures, when production from mixing is negligible.

\medskip
\textbf{\textit{Evolution of Sterile Neutrino DM.}}---
\label{sec:production}
The time-evolution of the sterile neutrino distribution function \( f_s(p,t) \) can be described by a semi-classical Boltzmann equation derived from the quantum kinetic equation (QKE)~\cite{Dodelson:1993je, Abazajian:2001nj}. This equation serves as a classical approximation to the full QKE framework, incorporating quantum corrections through an effective active--sterile transition probability. The approximation remains valid in the small mixing angle limit~\cite{Bell:1998ds, Kishimoto:2008ic, Johns:2019hjl}.

In the presence of active--sterile NSI, the Boltzmann equation is modified to
\begin{align}
    \label{eqn:modified_boltzmann_equation}
    \frac{\partial f_s(t, p)}{\partial t} - H p \frac{\partial f_s(t, p)}{\partial p} = \mathcal{C}_{\nu_a \leftrightarrow \nu_s} + \mathcal{C}_{\nu_a \nu_a \leftrightarrow \nu_s \nu_s},
\end{align}
where $H(t)$ is the Hubble rate. The first collision term on the right-hand side describes sterile neutrino production via active--sterile oscillations. It depends on both the total interaction rate \( \Gamma_{\rm tot} = \Gamma_{\mathrm{SM}} + \Gamma_{\mathrm{NSI}} \) and effective potential \( \mathcal{V}_{\mathrm{tot}} = \mathcal{V}_{\rm SM} + \mathcal{V}_{\mathrm{NSI}} \), which include contributions from the SM as well as NSI. The second collision term in Eq.~\eqref{eqn:modified_boltzmann_equation} accounts for number-changing interactions induced by NSI.

The SM contribution to the thermally averaged scattering rate of \( \nu_a \), arising from interactions with charged leptons, is given by~\cite{Abazajian:2001nj, Merle:2015vzu}
\begin{align}
    \Gamma_{\mathrm{SM}} (p) = C_a(T)\, G_F^2\, p\, T^4 \, ,
\end{align}
where \( G_F \) the Fermi constant and  \( C_a(T) \) is a temperature-dependent function defined in Refs.~\cite{Abazajian:2001nj, Merle:2015vzu, Asaka:2006nq}. 
The quantity \( \Gamma_{\mathrm{NSI}} \) denotes the interaction rate arising from additional scattering channels induced by NSI. In this scenario, the only relevant process is \( \nu_a \nu_s \leftrightarrow \nu_a \nu_s \), which contributes to the interaction rates of both active and sterile neutrinos. Specifically, this rate is given by the sum
\begin{align}
\Gamma_{\mathrm{NSI}} = \Gamma_{\nu_a \nu_s \leftrightarrow \nu_a \nu_s}^{(a)} + \Gamma_{\nu_a \nu_s \leftrightarrow \nu_a \nu_s}^{(s)} \, .
\end{align}
In the heavy-mediator limit \( m_\phi^2 \gg p T  \), these rates simplify  to (see {\it Supplemental Material} for the derivation)
\begin{align}
  \Gamma_{\nu_a \nu_s \leftrightarrow \nu_a \nu_s}^{(i)}(p) \simeq \frac{7 \pi y_{as}^4}{108\, m_\phi^4}\, p\,  \times 
\begin{cases}
\alpha T_s^4 & \text{for } i = a \\
T^4 & \text{for } i = s 
\end{cases} \, .
\end{align}

The SM effective potential for $\nu_a$ can be expressed as~\cite{Notzold:1987ik, DOlivo:1992lwg, Abazajian:2005gj} 
\begin{align}
\label{Eq:VT_SM}
\mathcal{V}_{\rm SM} (p) =\ & 
\frac{2 \sqrt{2} \zeta(3)\, G_F\, \eta_B\, T^3}{4 \pi^2}- \frac{8\sqrt{2}\, G_F\, p}{3 m_Z^2}  
\left(\rho_{\nu_a} +\rho _{\bar\nu_a} \right) \nonumber \\
& - \frac{8\sqrt{2}\, G_F\, p}{3 m_W^2}  
\left(\rho_a + \rho_{\bar a} \right)\, , 
\end{align}
where the first term is the baryonic contribution and the rest are the thermal contributions, with \( \rho_a, \rho_{\nu_a} \) being the energy densities of corresponding charged leptons and active neutrinos, respectively.
The NSI contribution to the thermal potential arises from the self-energy diagrams shown in Figs.~\ref{fig:selfenergy1} and \ref{fig:selfenergy2}, and is given by the difference between the active and sterile neutrino effective potentials~\cite{Dasgupta:2013zpn, Astros:2023xhe}:
\begin{align}
    \mathcal{V}_{\mathrm{NSI}} = \mathcal{V}_a - \mathcal{V}_s.
\end{align}
Assuming the sterile neutrino mass is negligible compared to the relevant temperatures, the thermal potential for a neutrino species \( \nu_i \) (with \( i = a \) or \( s \)) induced by the other species \( \nu_j \) (with \( j \neq i \)) in the heavy-mediator limit  takes the form~\cite{Astros:2023xhe}
\begin{align}
\mathcal{V}_i(p) \simeq - \frac{7\pi^2\, y_{as}^2\, }{45\, m_\phi^4} p\, \times
\begin{cases}
\alpha T_s^4 & \text{for } i = a \\
T^4 & \text{for } i = s
\end{cases} \, .
\end{align}

In our framework, sterile neutrinos can also be produced directly via the number-changing process \( \nu_a \nu_a \to \nu_s \nu_s \), as shown in Fig.~\ref{fig:numberchanging}. This contribution is captured by the second collision term in Eq.~(\ref{eqn:modified_boltzmann_equation}). At early times, when production from the modified DW mechanism is negligible, these interactions play a crucial role in initiating the sterile neutrino population and driving DM production. For a heavy mediator, the corresponding rates simplify to (see {\it Supplemental Material} for details)
\begin{align}
\Gamma_{\nu_i \nu_i \rightarrow \nu_j \nu_j}(p ) \simeq \frac{7 \pi y_{as}^4\,  }{216\, m_\phi^4} p \times 
\begin{cases}
T^4 & \text{for } i = a,\; j = s \\
\alpha T_s^4 & \text{for } i = s,\; j = a
\end{cases} \, .
\end{align}

\begin{figure}[t!]
     \centering
        \includegraphics[width=0.45\textwidth]{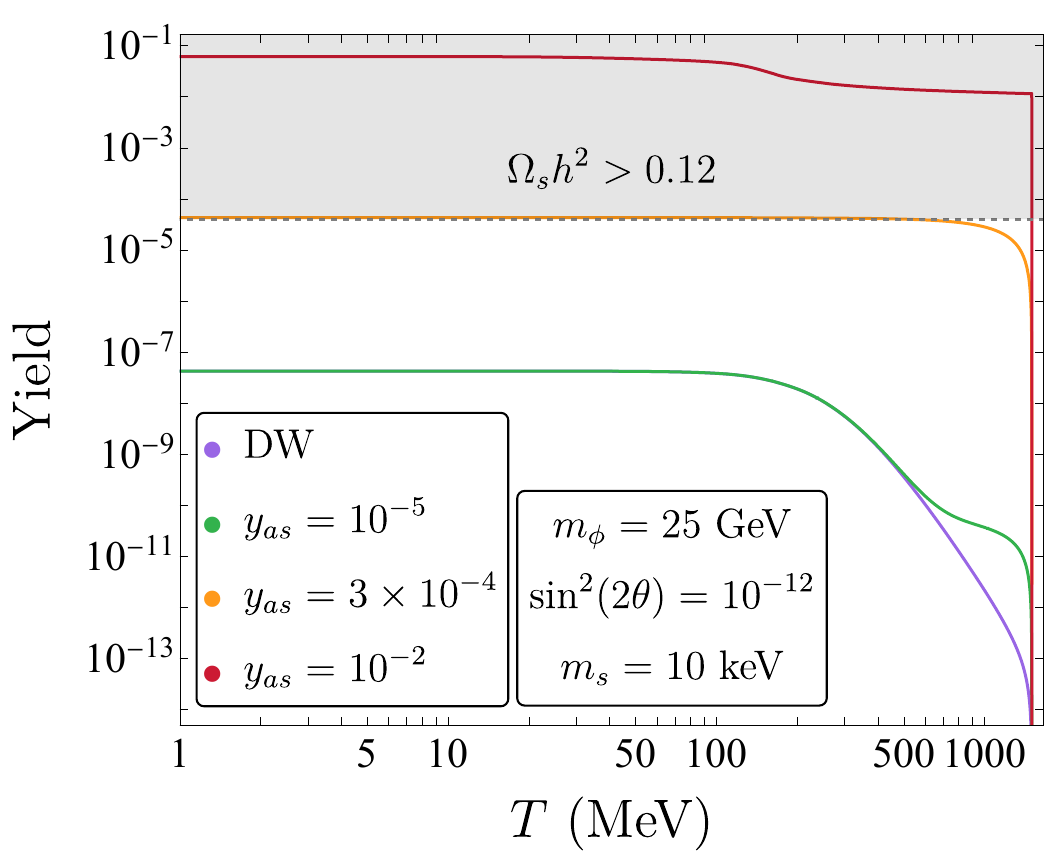}
\caption{\justifying 
Evolution of the comoving yield for  sterile neutrino DM production for the chosen  BP  as a  function of the SM temperature \( T \). 
}
\label{fig:yield}
\end{figure}
We solve the Boltzmann equation~\eqref{eqn:modified_boltzmann_equation} by first rewriting it in terms of the evolution equations for the comoving number density \( n_s \) and energy density \( \rho_s \) of sterile neutrinos. These equations are then solved numerically from an initial temperature of $T_i = 1.5\,\mathrm{GeV}$ up to the active neutrino decoupling temperature, \( T \approx 1\,\mathrm{MeV} \).\footnote{Our final result is insensitive to the choice of the initial temperature, as long as it is well below the mediator mass and well above the neutrino decoupling temperature.}
In Fig.~\ref{fig:yield}, we show the evolution of the sterile neutrino comoving yield 
for a representative benchmark point (BP) with \( m_s = 10\,\text{keV}, \,\sin^2 2\theta = 10^{-12},\, m_\phi = 25\,\text{GeV} \), shown by the red star in Fig.~\ref{fig:param_space}, which satisfies current astrophysical constraints and lies within the projected reach of future $X$-ray telescopes. We notice the following features:
(i) For very small couplings (e.g., \( y_{as} \lesssim  10^{-5} \)), number-changing interactions are only briefly active and quickly become inefficient. In this regime, modifications to the DW mechanism from the additional scatterings and thermal potentials are negligible, and the yield closely tracks the DW baseline. 
(ii) For slightly larger couplings (e.g., \( y_{as} \sim 10^{-4} \)), number-changing processes remain efficient for a longer duration, sustaining production down to \( T \sim 100\,\mathrm{MeV} \). As a result, the yield exceeds the DW curve before freezing in. After number-changing production ceases, the spectrum gradually returns to a DW-like shape.  
(iii) For large couplings (e.g., \( y_{as} \gtrsim  10^{-2} \)), number-changing interactions remain efficient throughout the entire production window, keeping the sterile sector in thermal contact with the active neutrinos. The yield quickly rises to the overproduction region (grey shaded), well above the DW line. 
This scenario is cosmologically excluded as it would indicate overproduction of DM beyond the observed value \( \Omega_s h^2 = 0.120\pm 0.001 \)~\cite{Planck:2018vyg}, as well as of extra number of relativistic degrees of freedom at BBN.

Note that the final relic abundance is determined by the suppression factor \( \alpha \) in Eq.~\eqref{eq:dist}  while the effective temperature \( T_s \) controls the spectral shape of the sterile neutrino population. The evolution of these quantities is given in the 
{\it Supplemental Material}. 

\textbf{\textit{Results.}}---
\label{sec:results}
Figure~\ref{fig:param_space} presents the viable parameter space in the plane of sterile neutrino mass and active–sterile mixing angle for a fixed value of $y_{as}=3 \times 10^{-4}$. We find that the presence of NSI significantly enlarges the allowed region beyond the traditional DW scenario. In particular, viable production can now happen at  much smaller mixing angles and even remains efficient in the limit of vanishing mixing. This behavior differs substantially from both the standard DW mechanism and earlier works on both active and sterile neutrino self-interaction scenarios \cite{DeGouvea:2019wpf, Kelly:2020pcy, Benso:2021hhh,Bringmann:2022aim, Johns:2019cwc, Astros:2023xhe}, where underproduction necessarily occurs at small mixing due to the lack of efficient production channels.

Moreover, in the previously discussed active-active and sterile-sterile NSI scenarios, for heavy mediators, a substantial departure from DW typically requires couplings \( \gtrsim 10^{-2} \)~\cite{Astros:2023xhe, DeGouvea:2019wpf}. However, in our case, such large couplings lead to overproduction of sterile neutrinos. In fact, efficient production can be achieved at smaller couplings without relying on active–sterile mixing. Figure~\ref{fig:param_space} indicates a transition in the parameter space: for a given coupling and mediator mass, there exists a minimum sterile neutrino mass above which number-changing interactions alone can generate the observed relic abundance irrespective of the mixing angle. As the mediator mass increases, the resulting abundance approaches that of the standard DW scenario, as expected in the heavy mediator regime.

\begin{figure}[t!]
    \centering
    \includegraphics[width=0.5
    \textwidth]{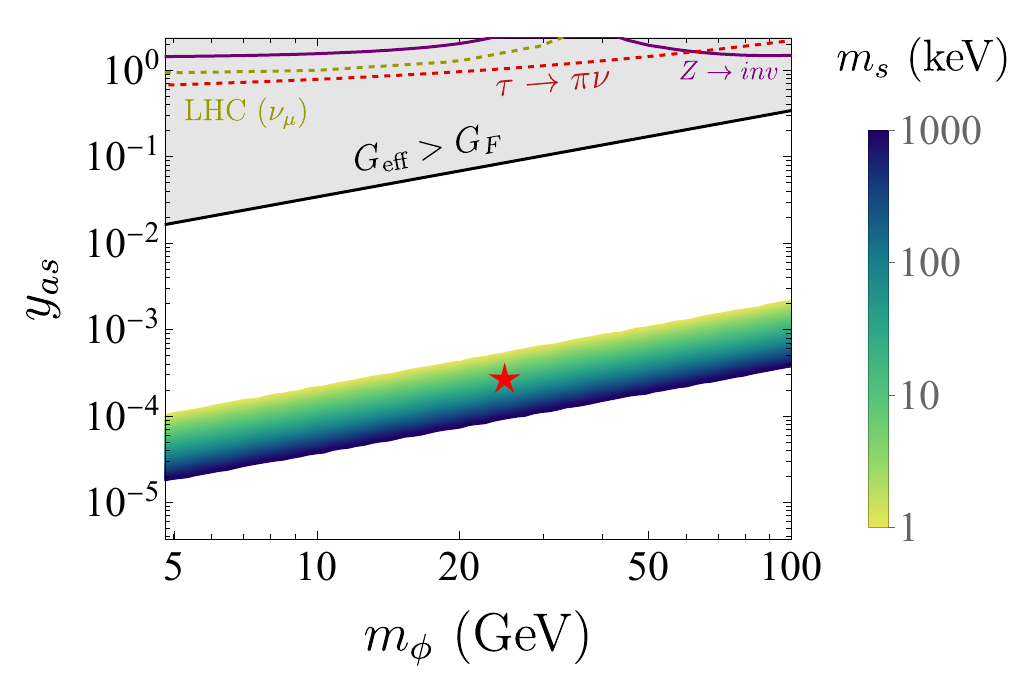}
     \captionsetup{justification=justified, singlelinecheck=false}
     \caption{\justifying 
Mediator mass versus coupling strength for allowed values of mixing angle. The colored band denotes the region with the correct DM abundance (\( \Omega_s h^2 = 0.12 \)). The laboratory constraints shown here are all within the cosmologically disfavored regime with $G_{\rm eff}>G_F$ (grey shaded). The red star corresponds to the BP shown in Fig.~\ref{fig:param_space} that gives the correct yield in Fig.~\ref{fig:yield}.}
\label{fig:param_space_y_mphi_nomixing}
\end{figure}

The viable parameter space now extends into regions where sterile neutrino DM can be produced efficiently without violating existing astrophysical constraints from $X$-ray telescopes such as Chandra~\cite{Hofmann:2016urz}, XMM-Newton~\cite{Borriello:2011un}, NuSTAR~\cite{Krivonos:2024yvm}, and XRISM~\cite{Yin:2025xad}, as well as from gamma-ray telescopes like INTEGRAL/SPI~\cite{Calore:2022pks} at higher energies. We also include projected sensitivities from future missions such as ATHENA~\cite{Barret:2019qaw}, eROSITA~\cite{eROSITA:2012lfj}, and eXTP~\cite{Malyshev:2020hcc}. The blue circle in Fig.~\ref{fig:param_space} represents the reported detection of an unidentified $X$-ray line near 3.5 keV~\cite{Boyarsky:2014jta, Bulbul:2014sua, Cappelluti:2017ywp} which would correspond to a 7.1 keV sterile neutrino, but the existence of the 3.5 keV line signal remains debatable~\cite{Dessert:2023fen}. In addition to astrophysical probes, sterile neutrino DM coupling to electron flavor can also be tested via precision $\beta$-decay experiments, as shown by the dotted and dashed red lines in Fig.~\ref{fig:param_space} for TRISTAN~\cite{KATRIN:2018oow} and HUNTER~\cite{Martoff:2021vxp} projected sensitivities.

Furthermore, cosmological constraints from structure formation impose additional restrictions on warm DM candidates like sterile neutrinos, which, due to their relatively long free-streaming length, suppress the formation of small-scale structures, a feature that must remain consistent with observations of the matter power spectrum, Lyman-$\alpha$ and Milky-Way satellites~\cite{Wang:2017hof, DES:2020fxi,Zelko:2022tgf, Newton:2024jsy, Tan:2024cek}. Specifically, Lyman-$\alpha$ forest measurements impose a lower bound on the warm DM mass, 
which corresponds to a present-day root-mean-square velocity of $v_{\text{rms},\,\text{today}} \lesssim 16\,\text{m/s}$~\cite{Garzilli:2019qki}. From this, we can infer that the sterile neutrino mass should satisfy a lower bound of $m_s \gtrsim 8\,\text{keV}$, which is stronger than the generic lower bound of $m_s\gtrsim 2$~keV for any fermionic DM from phase-space density arguments based on observations of dwarf galaxies~\cite{Tremaine:1979we, Boyarsky:2008ju, Merle:2015vzu, Bezrukov:2024qwr}.

Fig.~\ref{fig:param_space_y_mphi_nomixing} shows the allowed parameter space in the plane of mediator mass \( m_\phi \) and coupling \( y_{as} \), assuming sterile neutrino production is dominated by NSI. These results hold for any mixing angle allowed by astrophysical bounds.  We find that sterile neutrinos with masses in the range 1–100~keV can account for the observed relic abundance even in the absence of mixing, for couplings of order \( y_{as} \sim 10^{-4}-10^{-3} \).
At smaller couplings (\( y_{as} \lesssim  10^{-5} \)), the number-changing process \( \nu_a \nu_a \rightarrow \nu_s \nu_s \) becomes inefficient, and the combined effect of scattering and oscillations is also insufficient to generate the observed relic abundance. The range of couplings considered in our analysis remains consistent with current experimental constraints. For larger values (\( y_{as} \gtrsim 10^{-3} \)), sterile neutrinos are typically overproduced unless the mediator is sufficiently heavy to suppress the production rate. The gray shaded regions in Fig.~\ref{fig:param_space_y_mphi_nomixing} indicate parameter choices where the effective neutrino coupling $G_{\rm eff}\equiv y_{as}^2/m_\phi^2$ is greater than $G_F$ and is therefore cosmologically excluded, because it would hinder neutrino decoupling at MeV-scale prior to BBN.  Notably, we find that laboratory constraints from \( Z \rightarrow \text{invisible} \) decays, tau decays and LHC searches~\cite{Brdar:2020nbj, Dev:2024twk} are already encompassed within the cosmologically disfavored region. 

The lowest mediator mass considered in Fig.~\ref{fig:param_space_y_mphi_nomixing} is 5 GeV, because below this, for our choice of the initial temperature $T_i$, we  enter a regime where the  mediator can become kinematically accessible in the thermal bath during the initial stage of DM production, potentially modifying the sterile sector evolution and enabling additional scattering processes involving on-shell \( \phi \) exchange. A detailed treatment of this `light' mediator regime is left for future work.

\textbf{\textit{Potential Impact.}}---The freeze-in mechanism discussed here for the efficient production of $\nu_s$ DM via the interaction term~\eqref{eq:Leff} provides a novel way to realize observable neutrino-DM interactions while being consistent with the DM relic density requirement -- a feat otherwise deemed extremely challenging, if not impossible, in the thermal freeze-out scenario~\cite{Dev:2025tdv}. This result could have major implications for the future observations of supernova neutrinos (either galactic or diffuse) at upcoming large-volume neutrino detectors like JUNO~\cite{JUNO:2015zny}, Hyper-K~\cite{Hyper-Kamiokande:2018ofw} and DUNE~\cite{DUNE:2020ypp}, as well as high-energy astrophysical neutrinos at neutrino telescopes like IceCube~\cite{IceCube:2016zyt} and KM3NeT~\cite{KM3Net:2016zxf}. In particular, it introduces an energy-dependent opacity via $\nu_a\nu_s\to \nu_a\nu_s$ scattering mediated by $\phi$ that would attenuate the astrophysical neutrino flux passing through a DM halo, with detectable effects in the observed neutrino energy spectrum~\cite{Dev:2025tdv, Ng:2014pca, Fiorillo:2020jvy, Fiorillo:2020zzj, Cline:2022qld, Ferrer:2022kei, Doring:2023vmk, Bertolez-Martinez:2025trs,Mondol:2025uuw,He:2025bex}.

The presence of keV sterile neutrinos can also influence the evolution of core-collapse supernovae~\cite{Shi:1993ee, Raffelt:2011nc, Arguelles:2016uwb, Suliga:2019bsq, Ray:2023gtu}, but after taking into account the dynamical feedback effect~\cite{Syvolap:2019dat,Suliga:2020vpz, Ray:2024jeu}, it did not show up in Fig.~\ref{fig:param_space}. However, all these analyses were done in the context of $\nu_a-\nu_s$ oscillations only. Including NSI effects could change the picture significantly. For instance, the NSI process \( \nu_a \nu_a \rightarrow \nu_s \nu_s \) mediated by $\phi$ can potentially affect the core-collapse supernova dynamics by depleting the active neutrino population, thereby reducing the neutrino pressure and altering the cooling rate of the proto-neutron star. A detailed treatment of this effect is beyond the scope of the present work, but recent results from Refs.~\cite{Manzari:2023gkt, Cappiello:2025tws}  show that the ensuing constraints from SN1987A might become relevant in the parameter space shown in   Fig.~\ref{fig:param_space_y_mphi_nomixing}.

In addition, the $\nu_a\nu_a\to \nu_s\nu_s$ process would reduce the number of active neutrinos escaping from core-collapse supernova events throughout the cosmic history. This leads to a suppressed diffuse supernova neutrino background (DSNB) flux, as compared to the standard expectation ~\cite{Jeong:2018yts,DeGouvea:2020ang,Balantekin:2023jlg,Chauhan:2025hoz}. Given the recent hint  of a potential DSNB detection in Super-K Gd run data~\cite{Harada:2024, Castelvecchi:2024}, this could become another important probe of the parameter space shown in Fig.~\ref{fig:param_space_y_mphi_nomixing}. 

The NSI considered here could also have interesting cosmological implications. In particular, the $\nu_a\nu_a\to \nu_s\nu_s$ process with sub-MeV sterile neutrinos serves as an efficient neutrino cooling channel after the standard neutrino decoupling~\cite{Benso:2024qrg}. This could in fact mimic self-interacting neutrinos and resolve some persistent tensions in current  cosmological datasets~\cite{Das:2025asx}. 

UV-complete models of the effective interaction term~\eqref{eq:Leff} offer additional interesting phenomenology in laboratory experiments. For instance, the model proposed in Ref.~\cite{Dutta:2020scq} was originally motivated by the MiniBooNE, KOTO and muon $g-2$ anomalies. Although the latter two anomalies have since disappeared, the MiniBooNE excess still remains unresolved. 

In conclusion, our new mechanism for the freeze-in production of sterile neutrino DM has broad phenomenological implications. 

\acknowledgments
\textbf{\textit{Acknowledgments.}}---
\label{sec:acknowledgments}
We thank Maria Dias Astros and Stefan Vogl for useful discussion and correspondence on Ref.~\cite{Astros:2023xhe}, and Anna Suliga for useful discussion on the supernova constraints. We also thank the organizers of `Particle Physics on the Plains' (PPP) 2023 at University of Kansas, where this work was initiated. PSBD thanks the Mitchell Institute at Texas A\&M for local hospitality during the final stages of this work. The work of PSBD, JT and AUR was partly supported by the U.S. Department of Energy under grant No.~DE-SC0017987. PSBD was also partly supported by a Humboldt Fellowship from the Alexander von Humboldt Foundation. The work of BD is supported by the U.S. DOE
Grant~DE-SC0010813. 
SG acknowledges the J.C. Bose Fellowship (JCB/2020/000011) by the Anusandhan National Research Foundation, and Department of Space, Govt. of India. She also acknowledges Northwestern University (NU), where
a part of this work was done, for hospitality and Fulbright-Nehru Academic and Professional
Excellence fellowship for funding the visit to NU.
\bibliographystyle{apsrev4-2}
\bibliography{bib}
\appendix 
\onecolumngrid
\bigskip
\hrule
\bigskip
\begin{center}
{\bf \large Supplemental Material}
\end{center}

\setcounter{equation}{0}
\setcounter{figure}{0}
\setcounter{table}{0}
\makeatletter
\renewcommand{\theequation}{S\arabic{equation}}
\renewcommand{\thefigure}{S\arabic{figure}}
\renewcommand{\thetable}{S\arabic{table}}


\section{Collision terms}
\label{app:numberchanging}
The collision terms appearing in the Boltzmann equation (3) 
are given explicitly below \cite{Abazajian:2001nj}: 
\begin{align}
\label{eqn:collisionterm-oscillation}
    \mathcal{C}_{\nu_a \leftrightarrow \nu_s}(p) = & \frac{\Gamma_{\rm tot}}{4} \frac{\omega^2(p) \sin^2(2\theta)}{\omega^2(p) \sin^2(2\theta) + D^2(p) + [\omega(p) \cos(2\theta) - \mathcal{V}_{\mathrm{tot}}]^2} \left[f_a(t, p) - f_s(t, p)\right] \, , \\
\label{eqn:collisionterm-numberchanging}
    \mathcal{C}_{\nu_a \nu_a \leftrightarrow \nu_s \nu_s} (p) = & \frac{1}{2 E} \int \prod_{i=1}^{3} d\Pi_i \, (2\pi)^4 \delta^{(4)}(\mathbf{P}_1 + \mathbf{P}_2 - \mathbf{P}_3 - \mathbf{P})
\big(\, |\mathcal{M}_{\nu_a \nu_a \rightarrow \nu_s \nu_s}|^2 f_a(p_1)\, f_a(p_2)\, [1 \mp f_s(p_3)]\, [1 \mp f_s(p)] \nonumber \\
& \qquad \qquad - |\mathcal{M}_{\nu_s \nu_s \rightarrow \nu_a \nu_a}|^2 f_s(p_3)\, f_s(p)\, [1 \mp f_a(p_1)]\, [1 \mp f_a(p_2)] \,\big) \, .
\end{align}
Here, \( d\Pi_i \) denote the Lorentz-invariant phase space elements, the boldface $\mathbf{P}_i$'s are the 4-momenta, $p_i$'s are the magnitudes of the 3-momenta, and \( \omega(p) = \frac{m_s^2 - m_{\nu_a}^2 }{2p} \simeq \frac{m_s^2}{2p} \) is the vacuum oscillation frequency for \( m_s \gg m_{\nu_a} \). The total interaction rate \( \Gamma_{\rm tot} = \Gamma_{\mathrm{SM}} + \Gamma_{\mathrm{NSI}} \) and effective potential \( \mathcal{V}_{\mathrm{tot}} = \mathcal{V}_{\rm SM} + \mathcal{V}_{\mathrm{NSI}} \) incorporate both the SM contributions and modifications from NSI. The term \( D(p) = \Gamma_{\rm tot}/2 \) corresponds to the quantum damping rate, which accounts for decoherence effects in neutrino propagation.

The resonance condition for modified DW production is given by \cite{Astros:2023xhe}
\begin{align}
    \omega(p) \cos(2\theta) - \mathcal{V}_{\mathrm{tot}} = 0,
\end{align}
which yields two distinct resonance scenarios: a standard MSW-like resonance when \( \omega(p) \cos(2\theta) = \mathcal{V}_{\mathrm{tot}} \), and a second case where the total effective potential vanishes, \( \mathcal{V}_{\mathrm{tot}} = 0 \). These conditions typically require large fine-tuned couplings of order \( y_{as} \gtrsim \mathcal{O}(10^{-2}) \), which would lead to excessive production of sterile neutrinos through number-changing processes present in our NSI framework. For this reason, we do not consider resonance effects in this work.

We can simplify Eq.~(\ref{eqn:collisionterm-numberchanging}) by noting that \( |\mathcal{M}_{\nu_s \nu_s \rightarrow \nu_a \nu_a}|^2 = |\mathcal{M}_{\nu_a \nu_a \rightarrow \nu_s \nu_s}|^2 \equiv |\mathcal{M}|^2 \) due to crossing symmetry, and by neglecting Pauli blocking and Bose enhancement effects (i.e., \( 1 \mp f \approx 1 \)). Under these assumptions, the collision term reduces to:
\begin{align}
  \mathcal{C}_{\nu_a \nu_a \leftrightarrow \nu_s \nu_s} (p) = \frac{1}{2E} \int \prod_{i=1}^{3} d\Pi_i \, (2\pi)^4 \delta^{(4)}(\mathbf{P}_1 + \mathbf{P}_2 - \mathbf{P}_3 - \mathbf{P}) 
\, |\mathcal{M}|^2 \left[ f_a(p_1) f_a(p_2) - f_s(p_3) f_s(p) \right].
\end{align}
The zeroth and first moments of Eqs.~(\ref{eqn:collisionterm-oscillation}, \ref{eqn:collisionterm-numberchanging}):
\begin{equation}
\mathrm{C}_{n_s} = \int \frac{d^3 p}{(2\pi)^3} \,   \big( \mathcal{C}_{\nu_a \leftrightarrow \nu_s}(p) +\mathcal{C}_{\nu_a \nu_a \leftrightarrow \nu_s \nu_s} (p)\big), \qquad
\mathrm{C}_{\rho_s} = \int \frac{d^3 p}{(2\pi)^3} \, p \,   \big( \mathcal{C}_{\nu_a \leftrightarrow \nu_s}(p) +\mathcal{C}_{\nu_a \nu_a \leftrightarrow \nu_s \nu_s} (p)\big),
\end{equation}
 enter the integrated Boltzmann equations governing the evolution of the sterile neutrino number and energy densities:
\begin{align}
 \dot{n}_s + 3H n_s = \mathrm{C}_{n_s}, \qquad \dot{\rho}_s + 4H \rho_s = \mathrm{C}_{\rho_s}.
 \label{eq:Boltz_int}
\end{align}
The contribution from the number-changing process $\nu_a \nu_a \leftrightarrow \nu_s \nu_s$ can be recast in terms of the interaction rates as
\begin{align}
\mathrm{C}_{\nu_a \nu_a \leftrightarrow \nu_s \nu_s}^{(n)} = & \int \frac{d^3 p}{(2\pi)^3} \,   \mathcal{C}_{\nu_a \nu_a \leftrightarrow \nu_s \nu_s} (p) = \int \frac{d^3 p}{(2\pi)^3} \big( f_a(p)\, \Gamma_{\nu_a \nu_a \rightarrow \nu_s \nu_s}(p) - f_s(p)\, \Gamma_{\nu_s \nu_s \rightarrow \nu_a \nu_a}(p) \big), \nonumber \\
\mathrm{C}_{\nu_a \nu_a \leftrightarrow \nu_s \nu_s}^{(\rho)} = &   \int \frac{d^3 p}{(2\pi)^3} \, p \,  \mathcal{C}_{\nu_a \nu_a \leftrightarrow \nu_s \nu_s} (p) =\int \frac{d^3 p}{(2\pi)^3}\, p\,\big( f_a(p)\, \Gamma_{\nu_a \nu_a \rightarrow \nu_s \nu_s}(p) - f_s(p)\, \Gamma_{\nu_s \nu_s \rightarrow \nu_a \nu_a}(p) \big) \, .
\end{align}

\section{Interaction rates}
The general form for the interaction rate of a neutrino species \( \nu_i \) in a bath of \( \nu_j \) is \cite{Kelly:2020pcy}
\begin{align}
  \Gamma_{\nu_i \nu_j \, \leftrightarrow \, {\rm final}}^{(i)}(p) 
  = \int \frac{d^3 k}{(2\pi)^3} \, v_{\text{M\o ller}} \, f_j(k) \, \sigma_{ij\, \leftrightarrow \, {\rm final}}(p, k),
\end{align}
where the Møller velocity between the incident and target neutrinos is defined as~\cite{Cannoni:2013bza}
\begin{align}
v_\text{Møller} = \sqrt{\left( \vec{v}_{\rm in} - \vec{v}_{\rm tar} \right)^2 - (\vec{v}_{\rm in} \times \vec{v}_{\rm tar})^2 },
\end{align}
and \( \sigma_{\nu_i \nu_j \leftrightarrow \text{final}} \) is the corresponding scattering cross section.
In the ultra-relativistic limit, this simplifies to \( v_{\text{M\o ller}} = 1 - \cos\vartheta \), where \( \vartheta \) is the angle between incoming and target momenta. The rate can then be expressed in terms of Mandelstam variable $s$ as \cite{Doring:2023vmk}
\begin{align}
  \Gamma_{\nu_i \nu_j \leftrightarrow \text{final}}^{(i)}(p) 
  = \frac{1}{16\pi^2 p^2} \int_0^\infty d k \, f_j(k) \int_0^{4pk} d s \, s \, \sigma_{\nu_i \nu_j \rightarrow \text{final}}(s).
  \label{eq:S10}
\end{align}

The distribution functions for the active and sterile neutrino species are taken to be
\begin{align}
f_a(p) = \frac{1}{e^{p/T} + 1}, \quad f_s(p) = \frac{\alpha}{e^{p/T_s} + 1},
\end{align}
as explained in the main text.

For the process \( \nu_a \nu_s \rightarrow \nu_a \nu_s \), both \( s \)-channel and \( u \)-channel diagrams contribute. The total spin-averaged squared matrix element is given by
\begin{equation}
\overline{|\mathcal{M}| ^2}_{\nu_a \nu_s \rightarrow \nu_a \nu_s }= \frac{16 y_{as}^4 \left( m_\phi^4 (s^2 + s t + t^2) + m_\phi^2 \left( \Gamma_\phi^2 (s^2 + s t + t^2) - 3 s t (s + t) \right) + 3 s^2 (s + t)^2 \right)}{(\Gamma_\phi^2 m_\phi^2 + (m_\phi^2 - s)^2) (\Gamma_\phi^2 m_\phi^2 + (m_\phi^2 + s + t)^2)} \, ,
\end{equation}
where $\Gamma_{\phi} = \frac{y_{as}^2}{4\pi} m_{\phi}$ is the decay width of the scalar mediator \( \phi \), and \( s \), \( t \), \( u \) are the Mandelstam variables. The corresponding cross section is:
\begin{align}
\sigma_{\nu_a \nu_s \rightarrow \nu_a \nu_s }(s) &= \frac{y_{as}^4}{\pi s^2 \left( \Gamma_\phi^2 m_\phi^2 + (m_\phi^2 - s)^2 \right)} 
\Bigg[ \left( 2 m_\phi^2 - 3 s \right) m_\phi^2 (m_\phi^2 - s)  
\log\left( \frac{m_\phi^2}{m_\phi^2 + s} \right) 
+ \frac{s \left( 2 m_\phi^6 - 4 m_\phi^4 s + m_\phi^2 s^2 + 3 s^3 \right)}{m_\phi^2 + s} \Bigg].
\end{align}
The rate expressions simplify in the following limits:
\begin{equation}
\Gamma_{\nu_a \nu_s \leftrightarrow \nu_a \nu_s}^{(s)}(p) \approx 
\begin{cases}
\displaystyle \frac{7\pi y_{as}^4\, p T^4}{108\, m_\phi^4} & \text{for } p T \ll m_\phi^2 \\[1.5ex]
\displaystyle \frac{y_{as}^4 T^2}{16\pi\, p} & \text{for } p T \gg m_\phi^2
\end{cases}
\quad ; \quad
\Gamma_{\nu_a \nu_s \leftrightarrow \nu_a \nu_s}^{(a)}(p) \approx 
\begin{cases}
\displaystyle \frac{7\pi \alpha y_{as}^4\, p T_s^4}{108\, m_\phi^4} & \text{for } p T_s \ll m_\phi^2 \\[1.5ex]
\displaystyle \frac{\alpha y_{as}^4 T_s^2}{16\pi\, p} & \text{for } p T_s \gg m_\phi^2
\end{cases}.
\end{equation}

For the process $\nu_a \nu_a \leftrightarrow \nu_s \nu_s$, both $t$-channel and $u$-channel diagrams contribute. The total spin-averaged squared matrix element is given by:
\begin{align}
\left| \mathcal{M}_{\nu_a \nu_a \leftrightarrow \nu_s \nu_s}\right|^2 &= \frac{16 y_{as}^4 \left(m_\phi^4 \left(s^2 + 3 s t + 3 t^2\right) + m_\phi^2 \left(s^2 \left(\Gamma_\phi^2 - t\right) - s t \left(t - 3 \Gamma_\phi^2\right) + 3 \Gamma_\phi^2 t^2\right) + t^2 (s + t)^2\right)}{\left(\Gamma_\phi^2 m_\phi^2 + \left(m_\phi^2 - t\right)^2\right) \left(\Gamma_\phi^2 m_\phi^2 + \left(m_\phi^2 + s + t\right)^2\right)} \, .
\end{align}
Assuming the narrow width limit \( \Gamma_\phi \ll m_\phi \) and negligible neutrino masses \( m_s,\, m_{\nu_a} \ll \sqrt{s} \), the total cross section simplifies to:

\begin{equation}
    \sigma_{\nu_a \nu_a \leftrightarrow \nu_s \nu_s} (s)=\frac{  y_{as}^4 \left(5 \,m_\phi^2 + 3 s\right) }{2 \pi s^2}\left[\frac{s}{m_\phi^2 + s} + \frac{2 \,m_\phi^2 }{\left(2 \,m_\phi^2 + s\right)}\log \left(\frac{m_\phi^2}{m_\phi^2 + s}\right)\right].
\end{equation}
In the limiting cases, the corresponding rates become
\begin{equation}
\Gamma_{\nu_a \nu_a \rightarrow \nu_s \nu_s}(p) \approx
\begin{cases}
\displaystyle \frac{7 \pi y_{as}^4\, p T^4}{216\, m_\phi^4} & \text{for } p T \ll m_\phi^2 \\[1.5ex]
\displaystyle \frac{y_{as}^4\, T^2}{32 \pi\, p} & \text{for } p T \gg m_\phi^2
\end{cases}
\quad ; \quad
\Gamma_{\nu_s \nu_s \rightarrow \nu_a \nu_a}(p) \approx
\begin{cases}
\displaystyle \frac{7 \pi \alpha y_{as}^4\, p T_s^4}{216\, m_\phi^4} & \text{for } p T_s \ll m_\phi^2 \\[1.5ex]
\displaystyle \frac{\alpha y_{as}^4\, T_s^2}{32 \pi\, p} & \text{for } p T_s \gg m_\phi^2
\end{cases}.
\end{equation}
\section{Comparison of the interaction rates}
\begin{figure}[t!]
    \centering
    \includegraphics[width=0.5\linewidth]{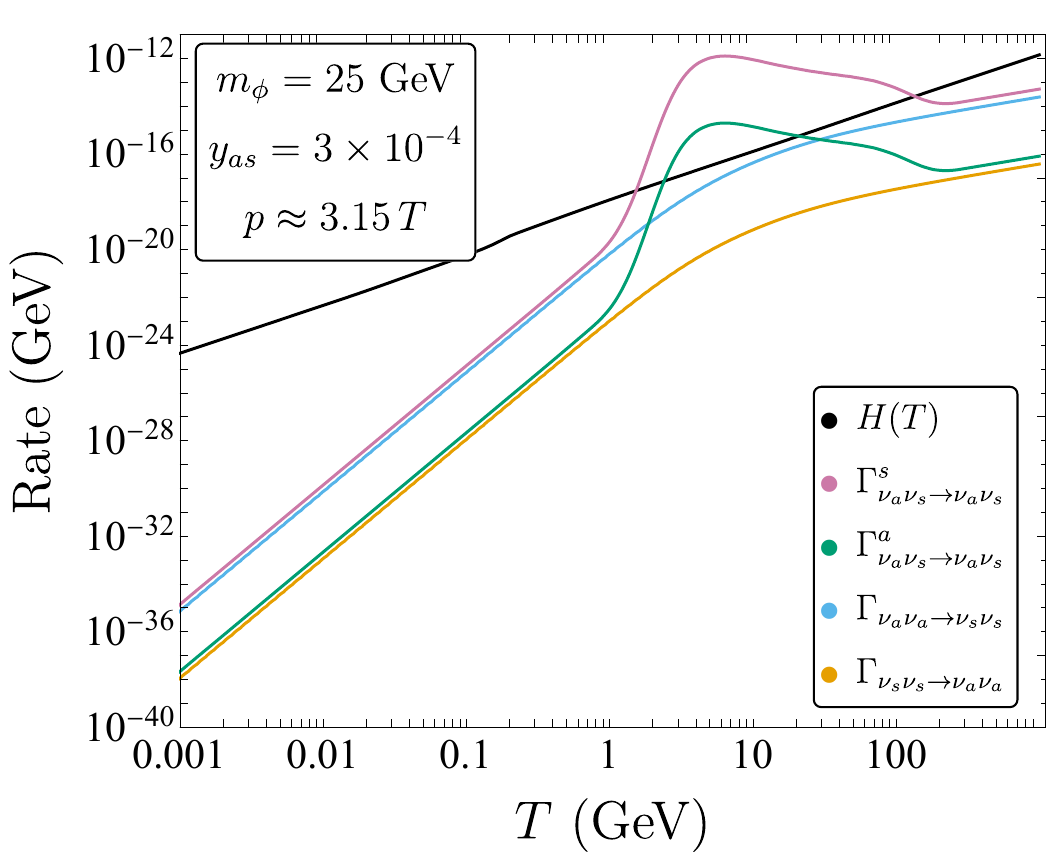}
    \caption{\justifying 
    Comparison of the relevant interaction rates to Hubble rate as functions of the photon temperature \( T \) for the BP (\( m_s = 10\,\mathrm{keV} \), \( y_{as} =   3 \times 10^{-4} \), \( m_\phi = 25\,\mathrm{GeV} \)). We have assumed $T_s \simeq T$ and $\alpha \sim 10^{-3}$ inferred from Fig. \ref{fig:alphaTS}.}
    \label{fig:HubblevsInteractionrate}
\end{figure}
We numerically compare the relevant interaction rates [cf. Eq.~\eqref{eq:S10}] to the Hubble expansion rate to assess chemical and kinetic equilibration for the benchmark point that reproduces the observed relic abundance (see Fig.~\ref{fig:yield}). As shown in Fig.~\ref{fig:HubblevsInteractionrate}, the interaction rates for $\nu_a\nu_a \leftrightarrow \nu_s\nu_s$ stay below $H(T)$ over the entire temperature range; consequently the sterile sector never attains chemical equilibrium and is produced via freeze-in, consistent with our assumption of zero initial abundance. By contrast, the elastic rate \( \nu_a\nu_s\!\leftrightarrow\!\nu_a\nu_s \) briefly exceeds \(H\) when $p T\sim m_\phi^2$,  efficiently establishing kinetic (but not chemical) equilibrium for a nonzero initial population of \( \nu_s \), but these rates do not affect the final yield of $\nu_s$. Accordingly, we model the sterile neutrinos distribution to be a modified Fermi-Dirac distribution, as detailed in the main text. Only for a large enough coupling \(y_{as}\gtrsim 10^{-3}\), the number-changing production rate \(\Gamma_{\nu_a\nu_a\to \nu_s\nu_s}\) can exceed the Hubble rate \(H\) near the resonance epoch, driving the sterile sector into chemical equilibrium with the active bath. In this regime, thermal equilibrium leads to over-abundance of DM (see Fig.~\ref{fig:yield}). At $T= 1\,$ MeV, all interaction rates satisfy $\Gamma \ll H$, so further sterile neutrino production is negligible.

\section{Evolution of $\alpha$ and $T_s$}

The parameters \( \alpha \) and \( T_s \) are defined by matching the sterile neutrino distribution \( f_s(p, T_s) = \alpha / (e^{p/T_s} + 1) \) to the number density and energy density. Assuming two spin degrees of freedom (\( g_s = 2 \)), we obtain

\begin{align}
n_s &\approx \frac{3 \zeta(3)}{2 \pi^2} \, \alpha\, T_s^3\,, \qquad \rho_s  \approx \frac{7 \pi^2}{120} \,\alpha\,  T_s^4
\end{align}
Solving for \( T_s \) and \( \alpha \), we get
\begin{align}
T_s &= \frac{180\, \zeta(3)}{7\, \pi^4}   \frac{\rho_s}{n_s} \,, \qquad
\alpha = \frac{2\pi^2}{3\zeta(3)}  \frac{n_s}{T_s^3} \,.
\end{align}

\begin{figure}[t!]
    \centering
    \includegraphics[width=0.46\linewidth]{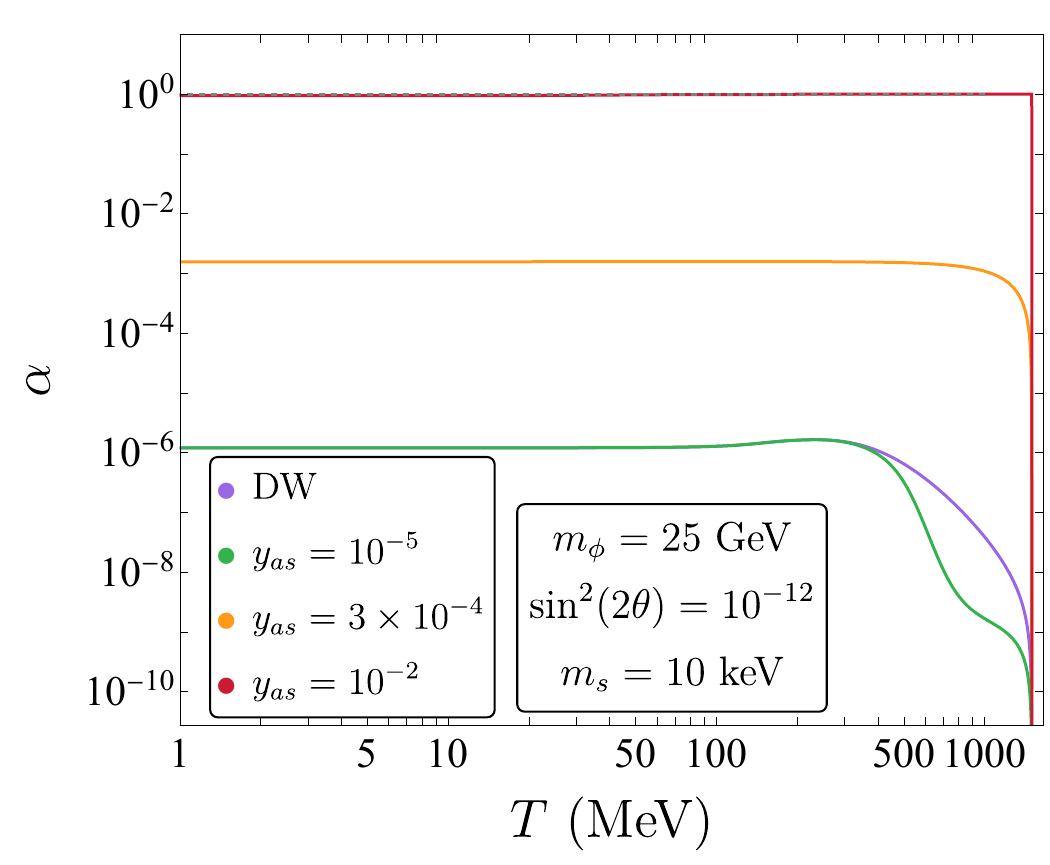}
    \hfill
    \includegraphics[width=0.45\linewidth]{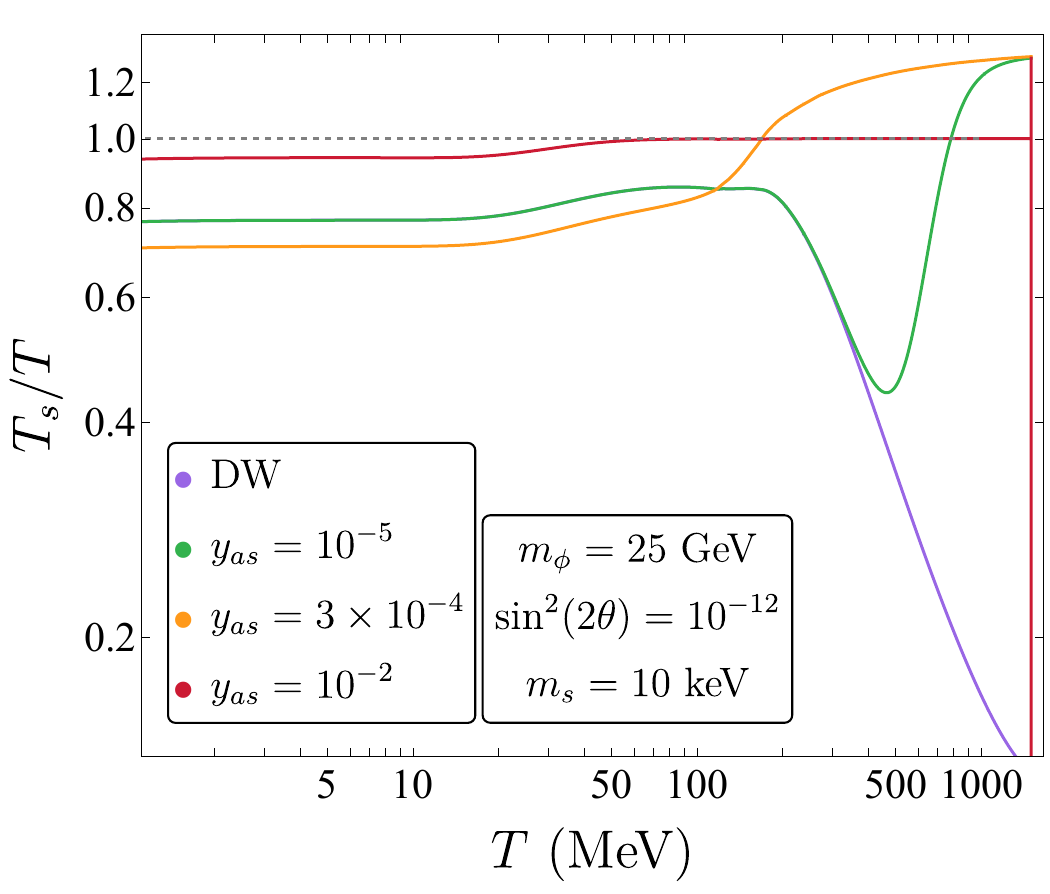}
    \caption{\justifying
    Evolution of the sterile neutrino distribution parameters \( \alpha \) and \( T_s \) for the BP (\( m_s = 10\,\mathrm{keV} \), \( \sin^2 2\theta =   10^{-12} \), \( m_\phi = 25\,\mathrm{GeV} \)) as functions of the photon temperature \( T \).}
    \label{fig:alphaTS}
\end{figure}

Figure~\ref{fig:alphaTS}  shows the evolution of the suppression factor $\alpha$ and the effective temperature $T_s$ with the SM temperature. From the left panel, it can be seen that  for very small couplings (e.g., \( y_{as} \lesssim  10^{-5} \)) the suppression factor remains small: \( \alpha \ll 1 \), resulting in an under-abundance of DM. 
For slightly larger couplings (e.g., \( y_{as} = 3 \times 10^{-4} \)), \( \alpha \) is higher as  compared to the lower-coupling case, ensuring  the correct relic abundance observed in Fig.~\ref{fig:yield}..
For even larger couplings (e.g., \( y_{as} =  10^{-2} \)), \( \alpha \approx 1 \) and $T_s \sim T $, signaling (near-)full thermalization of the sterile population leading to an over-abundance as was seen in Fig.~3. 
Note that in all cases, the extra interaction causes the effective temperature $T_s$ to be initially higher than the DW value, but it gradually converges to the DW case at lower temperatures for smaller Yukawa couplings as production from NSI ceases, while for large Yukawa couplings, $T_s \approx T$ implying that the sterile neutrino enters the thermal bath, which is therefore disfavored.  
The variation in  $T_s$ at neutrino decoupling across the parameter space, however, is modest in all cases considered, as can be seen from the right panel of Fig.~\ref{fig:alphaTS}. Together, the parameters \( \alpha \) and $T_s$ provide a complete characterization of sterile neutrino DM production in this framework. Note also that sterile neutrino contribution to the effective number of neutrino species at BBN is given by \( \Delta N_{\rm eff} \approx \alpha\,(T_s / T)^4 \). Thus, the current Planck limit of \( \Delta N_{\rm eff} < 0.214 \) at 95\% CL~\cite{Planck:2018vyg} sets an upper bound on $\alpha$, depending of the ratio $T_s/T$. 

\begin{figure}[t!]
    \centering
    \includegraphics[width=0.49\textwidth]{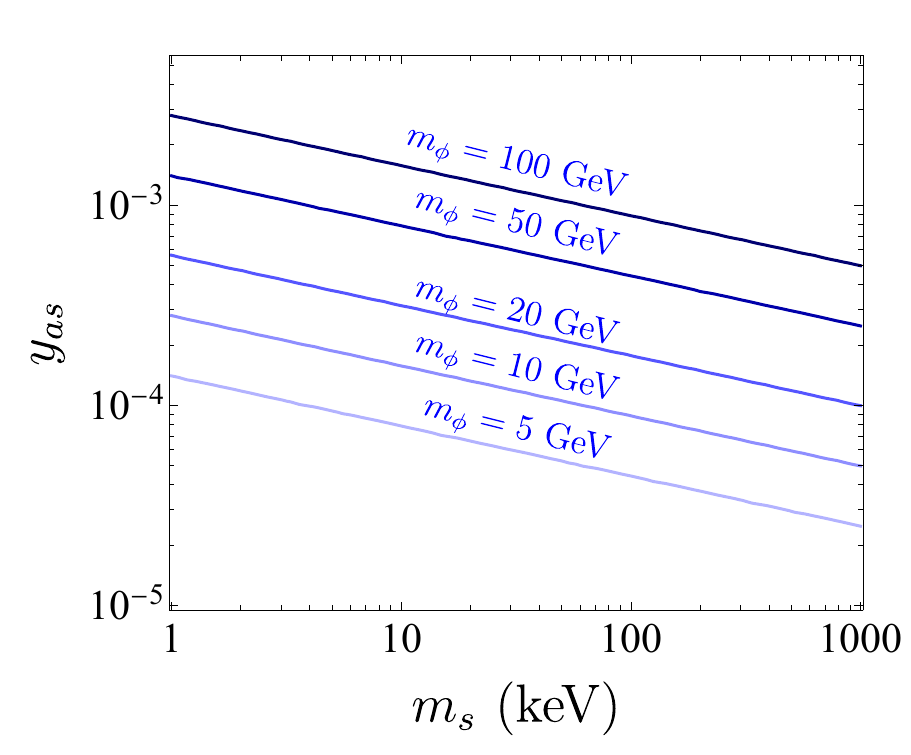} 
\includegraphics[width=0.49\textwidth]{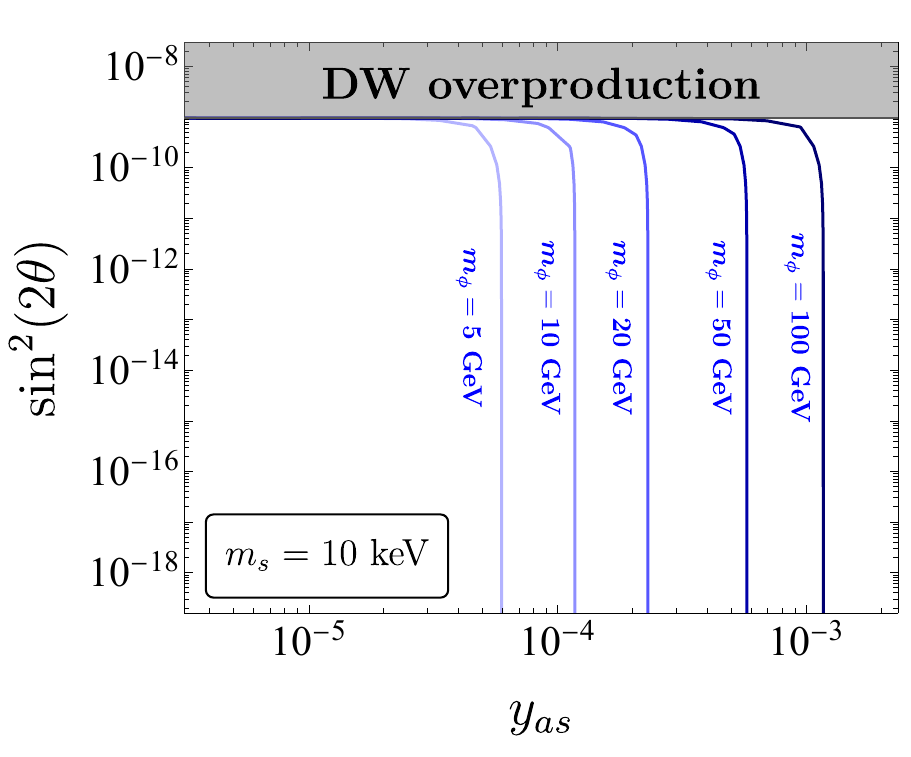}
        \caption{\justifying {\it Left:} Sterile neutrino mass vs. coupling strength parameter space for active-sterile mixing angle $\sin^2 2\theta = 0$. Blue contour lines indicate where the relic abundance satisfies $\Omega_s h^2 = 0.12$. {\it Right:}
         Parameter space of mixing angle versus coupling strength for a 10~keV sterile neutrino. Contours show values of \( \Omega_s h^2 = 0.12 \) for different mediator masses. The gray shaded region indicates where the DW mechanism alone leads to overproduction.}
        \label{fig:param_space_y_sinth}
\end{figure}

\section{Additional plots of the parameter space}

Figure~\ref{fig:param_space_y_sinth} provides complementary information to what has been shown in Figs.~1 and 4. 
The left panel of Fig.~\ref{fig:param_space_y_sinth} shows the variation of $y_{as}$ as a function of $m_s$ for different values of $m_\phi$ and for vanishing mixing angle. The correct relic density of $\nu_s$ is produced along the blue contours. We see that for larger (smaller) values of the mediator mass, we require larger (smaller) values of the Yukawa coupling to get the correct relic density. 

The right panel of Fig.~\ref{fig:param_space_y_sinth} shows the variation of $\sin^2(2\theta)$ as a function of $y_{as}$ for different values of $m_\phi$ and for a fixed DM mass of 10 keV. As in the left panel, the correct relic density is obtained along the blue contours. We see that depending on the mediator mass, the DM production is dominantly governed by the Yukawa coupling, irrespective of the mixing angle. But as the Yukawa coupling becomes small, we approach the DW limit, where the DM production is governed by the mixing angle. The gray shaded region on the top corresponds to large mixing angles for which even the DW mechanism leads to DM overproduction. 

\end{document}